\documentclass[prd,aps,12pt]{revtex4}
\usepackage{amsmath,amssymb} 
\usepackage[dvips]{graphicx} 
\usepackage{graphics}
\usepackage{epsfig}

\newcommand\ba{\begin{eqnarray}}
\newcommand\ea{\end{eqnarray}}
\newcommand\be{\begin{eqnarray}}
\newcommand\ee{\end{eqnarray}}
\newcommand\nn{\nonumber}

\begin{document}

\title{Final state emission radiative corrections to the process $e^+e^- \to \pi^+\pi^-(\gamma)$. Contribution to muon anomalous magnetic moment}
\author{A.~I.~Ahmadov$^{1,3}$~\footnote{E-mail: ahmadov@theor.jinr.ru},  E.~A.~Kuraev$^{1}$~\footnote{E-mail: kuraev@theor.jinr.ru},
M.~K.~Volkov$^{1}$~\footnote{E-mail: volkov@theor.jinr.ru},
O.~O.~Voskresenskaya$^{2}$~\footnote{E-mail: voskr@jinr.ru} and E.~V.~Zemlyanaya$^{2}$~\footnote{E-mail: elena@jinr.ru}}
\affiliation{$^{1}$ Bogoliubov Laboratory of Theoretical Physics,
Joint Institute for Nuclear Research, Dubna, 141980 Russia, \\
$^{2}$ Laboratory of Information Technologies,
Joint Institute for Nuclear Research, Dubna, 141980 Russia,  \\
$^{3}$ Institute of Physics, Azerbaijan
National Academy of Sciences, Baku, Azerbaijan
}

\date{\today}

\begin{abstract}
Analytic calculation are presented of the contribution to the anomalous magnetic moment of a muon
from the channels of annihilation of an electron-positron pair
to a pair of charged pi-mesons with radiative correction connected with the final state, as well as corrections to the lowest order kernel. The result with point-like
pi-meson assumptions is $a_{point}=a_{point}^{(1)}+\Delta a_{point}$,\,\, $a_{point}^{(1)}=7.0866 \cdot 10^{-9}; \,\, \Delta a_{point}= -2.4 \cdot 10^{-10} .$
Taking into account the pion form factor in the frames of the Nambu-Jona-Lasinio (NJL) approach leads to
$a_{NJL}=a_{NJL}^{(1)}+\Delta a_{NJL}$,\,\, $a_{NJL}^{(1)}= 5.48\cdot 10^{-8}; \,\, \Delta a_{NJL}=-3.43\cdot 10^{-9}$.
\end{abstract}

\maketitle

\section{General Formalism}

It is known \cite{Davier} that  about $73 \%$ of the contribution of hadrons to the
anomalous magnetic moment of the
muon, $a_\mu=(g-2)/2$ \cite{Brodsky1,Brodsky2}
\ba
a_\mu=\frac{1}{4\pi^3}\int\limits_{4m_\pi^2}^\infty d s \sigma^{e^+e^- \to \pi^+\pi^-}_B(s)K^{(1)}(\frac{s}{M^2}),
\ea
with $\sigma_B(s)$ being the total cross section in the Born approximation and
\ba
K^{(1)}(\frac{s}{M^2})=\int\limits_0^1d x\frac{x^2(1-x)}{x^2+\frac{s}{M^2}(1-x)},
\ea
with $M$ being the muon mass,
arises from taking into account the simplest process $e^+e^-\to \gamma^* \to \pi^+\pi^-$, whereas about
$60 \%$ of the error arises from the uncertainties associated with the pion pair production from the
mechanisms with intermediate states of the lightest vector meson $\rho,\omega$
\cite{Ven}.

It seems "natural" to use the result of experimental measuring of the cross section of the process $e^+e^-\to\pi^+\pi^-$.
But, unfortunately, the experimentally measured total cross section (omitting the effects of detection of the final particles)
includes the emission of both virtual and real photons by the initial electron and positron (ISE) and the final state emission (FSE),
and possibly, the interference of amplitudes of the emission of the initial and final particles. Assuming that the contribution of these interference terms to the total
cross section is zero (charge-blind setup), we remain with the problem of including such enhanced factors as the
form factor $F_\pi(s)$ of the charged pion in the timelike region and the delicate procedure of extracting the effects of the initial state emission (both photons and charged particles).
Only part of the radiative corrections to FSE connected with the final $\pi^+\pi^-$ can be included in the frames of one virtual photon polarization operator used above, since one implies $\sigma^{e^+e^- \to  hadrons}(s)=((4\pi\alpha)^2/s)Im\Pi(s)$.
The polarization operator is defined as a transverse part of the virtual photon self-energy tensor
$\Pi_{\mu\nu}(q)=(q_\mu q_\nu-q^2g_{\mu\nu})\Pi(q^2)$ and by applying the dispersion relation \cite{Krause,Brodsky1,Brodsky2}
\ba
\Pi(q^2)=-\frac{q^2}{\pi}\int\limits_{4m^2}^\infty\frac{d s}{s}\frac{Im\Pi(s)}{q^2-s},
\ea
where $m$ is the pion mass.

Replacing the Green function of the virtual photon in the one-loop vertex function by the one containing the polarization operator
\ba
-i\frac{g_{\mu\nu}}{q^2}\to -\frac{1}{\pi}\int\limits_{4m^2}^\infty\frac{d s}{s}Im\Pi(s)\frac{-ig_{\mu\nu}}{q^2-s},
\ea
one arrives at the known result of the lowest-order contribution to $a_\mu$ from the hadronic intermediate state
\ba
a_\mu^{(1)}=\frac{\alpha^2}{3\pi^2}\int\limits_{4m^2}^\infty d s\frac{R(s)K^{(1)}(s/M^2)}{s},\,\,\,\,
R(s) = \frac{3s}{4\pi\alpha^2}\sigma^{e^+e^- \to had.}(s),
\ea
and the lowest-order kernel is
\ba
K^{(1)}(s/M^2)=\int\limits_0^1 d x\frac{x^2(1-x)}{x^2+(1-x)(s/M^2)}.
\ea

The problem consists of removing from the experimentally measured cross section the radiative corrections associated with the initial electron-positron state, including the emission of virtual and real photon.
This procedure can be a source of errors and uncertainties.

One can include the pion form factor in the form of the replacment
\ba
Im\Pi(s) \to F_\pi^2(s) Im\Pi(s).
\ea
Below, we calculate the contribution to $a_\mu$ from the processes $e^+e^- \to \pi^+\pi^-$ and $e^+e^- \to \pi^+\pi^-(\gamma)$
assuming the pion to be a pointlike particle, taking into account the emission of virtual and real photons by the charged pions only.
To obtain the explicit formulas describing FSE is the motivation of our paper.

The differential (center-of-mass reference frame (cmf) is implied) and total cross sections of the process
\ba
e^+(p_+)+e^-(p_-) \to \pi^+(q_+)+\pi^-(q_-)
\ea
in the lowest-order of perturbation theory and with the assumption of point-like pion interaction with the virtual photon have the form
\ba
\frac{d\sigma}{d c}=\frac{\pi\alpha^2\beta^3}{4s}(1-c^2); \,\,\,\,
\sigma(s)=\frac{\pi\alpha^2\beta^3}{3s}, \,\,\,\,\beta=\sqrt{1-\frac{4m^2}{s}},
\ea
where $s=(p_++p_-)^2=4E^2$ is the square of the total energy $c=\cos\theta$, and $\theta$ is the angle between the directions of the initial electron and the negative charged pion in the cmf.
Inserting the explicit value of the total cross section we obtain
\ba
a_\mu^{(1)}=\frac{\alpha^2\rho^2}{6\pi^2}\int\limits_0^1d x x^2(1-x)\int\limits_0^1\frac{d\beta \beta^4}{4(1-x)+x^2\rho^2(1-\beta^2)}, \nn \\
\rho=\frac{M}{m}.
\ea
Numeric estimations give $a_\mu^{(1)}=7.08665\times 10^{-9}$.

In the next order of perturbation theory we must consider the contribution arising from the correction associated with the emission of virtual and real photons (soft and hard) by the final $\pi^+\pi^-$ state. It results in the replacement $\sigma(s) \to \sigma(s)(1+\delta(s))$.
Keeping in mind the correction to the kernel we obtain
\ba
a_\mu=\frac{1}{4\pi^3}\int\limits_{4m^2}^\infty d s\sigma(s)(1+\delta(s))[K^{(1)}(s/M^2)+\frac{\alpha}{\pi}K^{(2)}(s/M^2)].
\ea
The quantity $K^{(2)}(s/M^2)$ was computed in Ref. \cite{BR}. It is presented in the Appendix.
Radiative correction to the final state $\pi^+\pi^-$ will be considered below.

\section{Emission of virtual photons}

To start with a virtual correction, we find first the vertex function for scattering of the
charged pion in the external field.
Then we write it down in the annihilation channel and use it to calculate the relevant virtual
correction to the cross section.
The vertex function of the process $\pi^-(p_1)+\gamma^*(q) \to \pi^-(p_2)$ has the form
\ba
\Gamma_\mu=\frac{\alpha}{4\pi}\int\frac{N_\mu d k}{(k)(1)(2)}, (k)=k^2-\lambda^2;(1)=k^2-2p_1k; (2)=k^2-2p_2k, \nn \\
d k=\frac{d^4 k}{i\pi^2}, N_\mu=(p_1+p_2-2k)_\mu (2p_1-k)_\lambda(2p_2-k)^\lambda.
\ea
Writing $N_\mu$ as $N_\mu=(p_1+p_2-2k)_\mu[4p_1p_2+(1)+(2)-(k)]$ and performing the loop momenta integration,
we obtain for the unrenormalized vertex function
\ba
\Gamma^{un}_\mu=\frac{\alpha}{4\pi}(p_1+p_2)_\mu F^{un}(q^2), \nn \\
F^{un}(q^2)=(2m^2-q^2)\int\limits_0^1\frac{d x}{q_x^2}[\ln\frac{m^2}{\lambda^2}+\ln\frac{q_x^2}{m^2}-1]+3+\ln\frac{\Lambda^2}{m^2}, \nn \\
p_1^2=p_2^2=m^2, \,\,\,q_x=p_1x+p_2(1-x), \,\,\,q_x^2=m^2-x(1-x)q^2, \,\,\,q=p_2-p_1.
\ea
Here $\Lambda$ and $\lambda$ are the ultraviolet cutoff parameter and the fictitious photon mass.
The regularization consists in the construction $F(q^2)=F^{un}(q^2)-F^{un}(0)$. So we have
\ba
\Gamma_\mu=\frac{\alpha}{4\pi}(p_1+p_2)_\mu F(q^2), \nn \\
F(q^2)=(2m^2-q^2)\int\limits_0^1\frac{d x}{q_x^2}[\ln\frac{m^2}{\lambda^2}+\ln\frac{q_x^2}{m^2}-2]+4[1-\ln\frac{m}{\lambda}].
\ea
Introducing the new variable $(1-\theta)^2,\,\,\theta=-q^2/m^2$ and using
\ba
\int\limits_0^1\frac{d x}{q_x^2}=\frac{2\theta}{m^2(1-\theta^2)}\ln\frac{1}{\theta}, \nn \\
\int\limits_0^1\frac{d x}{q_x^2}\ln\frac{q_x^2}{m^2}=\frac{2\theta}{m^2(1-\theta^2)}[\frac{1}{2}\ln^2\theta-2\ln\theta \ln(1+\theta)-2Li_2(-\theta)-\frac{\pi^2}{6}],
\ea
we obtain
\ba
\Gamma_\mu(p_1,p_2)&=&\frac{\alpha}{\pi}(p_1+p_2)_\mu
\left[(\ln\frac{m}{\lambda}-1)(\frac{1+\theta^2}{1-\theta^2}-ln\frac{1}{\theta}-1)+
\right.
\nn\\
&+&\left.
\frac{1+\theta^2}{4(1-\theta^2)}[\ln^2\theta-4\ln\theta \ln(1+\theta)-4Li_2(-\theta)-\frac{\pi^2}{3}\right].
\ea
For the crossing channel $\gamma^*(q,\mu) \to \pi^-(q_-)+\pi^+(q_+)$ we use the substitutions \cite{BMR}
\ba
p_2\to q_-, \,\,\,p_1\to -q_+, \,\,\,\theta \to -x+i\epsilon, \,\,\,0<\epsilon<<1, \nn \\
x=\frac{1-\beta}{1+\beta}.
\ea
This quantity acquires the imaginary part for $s>4m^2$:
\ba
\Gamma_\mu&=&\frac{\alpha}{\pi}(q_--q_+)_\mu F(x),  \nn \\
F(x)&=&(\ln\frac{\lambda}{m}+1)(\frac{1+x^2}{1-x^2}\ln x+1+i\pi)+
\nn\\
&+&\frac{1+x^2}{4(1-x^2)}[\ln^2 x-\frac{4}{3}\pi^2-
4Li_2(x)-4\ln x\ln(1-x)+i\pi(2\ln x-4\ln(1-x))].\nn
\ea
Writing $Re F(x)$ as
\ba
Re F(x)=(-1+\frac{1+\beta^2}{2\beta}L)\ln\frac{\lambda}{m}+F_V,\nn \\
L=\ln\frac{1+\beta}{1-\beta},
\ea
we write down the relevant contribution to the total cross section as
\ba
\Delta_V\sigma(s)=\frac{2\alpha^3\beta^3}{3s}[(-1+\frac{1+\beta^2}{2\beta}L)\ln\frac{\lambda}{m} +F_V(\beta)],
\ea
with
\ba
F_V(\beta)=-1+\frac{1+\beta^2}{2\beta}[L-\frac{1}{4}L^2+\frac{1}{3}\pi^2+Li_2(\frac{1-\beta}{1+\beta})-L\ln\frac{2\beta}{1+\beta}].
\ea

\section{Emission of soft photon}

Consider now the contribution from the emission of the soft real photon channel. It has the form
\ba
\Delta_S\sigma=-\frac{\alpha}{4\pi^2}\sigma_B(s)\int^{'}\frac{d^3 k}{\omega}\biggl(\frac{q_-}{q_-k}-\frac{q_+}{q_+k}\biggr)^2,
\ea
where the prime means $\omega=\sqrt{\vec{k}^2+\lambda^2}<\Delta E$ and it is implied that $\Delta E<<E$.
Using the relations
\ba
\frac{\alpha}{4\pi^2}\int^{'}\frac{d^3 k}{\omega}\frac{m^2}{(q_-k)^2}=\frac{\alpha}{\pi}[\ln\frac{2\Delta E}{\lambda}-\frac{1}{2\beta}L], \qquad L=\ln\frac{1+\beta}{1-\beta},
\ea
\ba
\frac{\alpha}{4\pi^2}\int^{'}\frac{d^3 k}{\omega}\frac{(q_+q_-)}{(q_+k)(q_-k)}=\frac{\alpha}{\pi}\frac{1+\beta^2}{2\beta}
[\ln\frac{2\Delta E}{\lambda} L+J(\beta)],
\ea
with
\ba
J(\beta)=Li_2(-\beta)-Li_2(\beta)+Li_2(\frac{1+\beta}{2})-\frac{1}{4}L^2+\frac{1}{2}\ln^2(\frac{1+\beta}{2})-\frac{1}{12}\pi^2,
\ea
we obtain the contribution to the cross section
\ba
\Delta_S\sigma(s)=\frac{2\alpha^3\beta^3}{3s}[(-1+\frac{1+\beta^2}{2\beta}L)\ln\frac{2\Delta E}{\lambda} +F_S(\beta)], \nn \\
F_S(\beta)=-\frac{1}{2\beta L}+\frac{1+\beta^2}{2\beta}J(\beta).
\ea
Here the prime means $\omega < \Delta E$.

\section{Hard real photon emission}

Consider at least the contribution from the hard photon emission channel $\omega > \Delta E$.
The matrix element of this process has the form
\ba
M=\frac{(4\pi\alpha)^{3/2}}{s}J^\mu T_{\mu\nu}e(k)^\nu,
\ea
where $e(k)$ is the polarization vector of the photon, $J^\mu=\bar{v}(p_+)\gamma^\mu u(p_-)$ is the
current associated with the leptons, and
\ba
T_{\mu\nu}=\frac{1}{2(q_-k)}(2q_-+k)_\nu (Q+k)_\mu+\frac{1}{2(q_+k)}(-2q_+-k)_\nu (Q-k)_\mu-2g_{\mu\nu}.
\ea
It can be checked that this expression obeys the gauge invariance conditions $T_{\mu\nu}q^\mu=T_{\mu\nu}k^\nu=0$.
We use as well the relation \cite{QED}
\ba
\sum_{spin}\int |M|^2 d\Gamma_3=-\frac{1}{3}Tr\hat{p}_+\gamma^\mu\hat{p}_-\gamma^\nu (g_{\mu\nu}-q_\mu q_\nu/q^2) \int  I d\Gamma_3, \nn \\
I=T_{\rho\sigma}T^{\rho\sigma},
\ea
with $d\Gamma_3$ being the element of the phase space of the final particles,
\ba
d\Gamma_3=\frac{d^3q_-}{2E_-}\frac{d^3q_+}{2E_+}\frac{d^3 k}{2\omega}\frac{1}{(2\pi)^5}\delta^4(q-q_--q_+-k).
\ea
It can be written as
\ba
\frac{d^3q_-}{2E_-}\frac{d^3q_+}{2E_+}\frac{d^3 k}{2\omega}\frac{1}{2E_+(2\pi)^5}\delta(q_0-E_--E_+-\omega),
\ea
with $E_+=\sqrt{(\vec{q}_-+\vec{k})^2+m^2}=\sqrt{\omega^2+E_-^2+2\vec{q}_-\vec{k}}$.
Performing the integration over $\cos \theta$, where $\theta$ is the angle in the cmf between three-momenta of pion and photon, we obtain
\ba
d\Gamma_3=\frac{\pi^2 s}{4(2\pi)^5}d\nu d\nu_- d\nu_+\delta(\nu+\nu_-+\nu_+), \,\,\, \nu=\frac{2\omega}{q_0},\,\,\,\nu_-=\frac{2E_-}{q_0},\,\,\,\nu=\frac{2E_+}{q_0},\,\,\,q_0=2E.
\ea
In terms of energy fractions we obtain
\ba
I&=&8+\frac{2(1-\nu)}{1-\nu_+}+\frac{2(1-\nu)}{1-\nu_-}-\beta^2(1-\beta^2)[\frac{1}{(1-\nu_+)^2}+\frac{1}{(1-\nu_-)^2}]+
\nn\\
&+&\frac{2}{\nu}(\nu-\beta^2)(\nu-1-\beta^2)[\frac{1}{1-\nu_+}+\frac{1}{1-\nu_-}].
\ea
The domain of integration $D$ is
\ba
\frac{\Delta E}{E} <\nu< \beta^2; \,\,\,\nu+\nu_-+\nu_+=2; \nn \\
(1-\nu_-)(1-\nu_+)(1-\nu)>\frac{m^2\nu^2}{s}, \,\,\,\frac{\nu}{2}(1-R)<1-\nu_\pm<\frac{\nu}{2}(1+R), \,\,\, R=\sqrt{\frac{\beta^2-\nu}{1-\nu}}.
\ea
Performing the integration over $\nu_\pm$ we obtain
\ba
\int I d\nu_- d\nu_+\delta(2-\nu-\nu_--\nu_+)=4[2 R(\nu-\frac{\beta^2(1-\nu)}{\nu})+(\frac{\beta^2(1+\beta^2)}{\nu}-2\beta^2)\ln\frac{1+R}{1-R}]. \,\,\,\,\,
\ea
Performing further integration we use the substitution $t=R, \,\,0<t<t_m, \,\,t_m^2=\beta^2-(\Delta E/E)(1-\beta^2)$.
The corresponding contribution to the cross section is
\ba
\Delta_H\sigma(s)=\frac{2\alpha^3\beta^3}{3s}[(\frac{1+\beta^2}{2\beta}L-1)\ln\frac{E}{\Delta E}+F_H(\beta)],
\ea
with
\ba
F_H(\beta)= -\frac{1+\beta^2}{\beta}G(\beta)+\ln\frac{1-\beta^2}{4\beta^2}-\frac{1}{8\beta^3}(3+\beta^2)(1-\beta^2)L+\frac{3+7\beta^2}{4\beta^2},
\ea
and
\ba
G(\beta)=\int\limits_0^\beta\frac{d t}{1-t^2}\ln\frac{1-t^2}{\beta^2-t^2}=Li_2(\frac{1-\beta}{2})-Li_2(\frac{1+\beta}{2})+Li_2(1+\beta)-Li_2(1-\beta).
\ea

The total contribution does not depend on the "photon mass" $\lambda$ or on the auxiliary parameter $\Delta E$:
\ba
\Delta\sigma=\frac{2\alpha^3\beta^3}{3s}\biggl[\frac{1}{2}\biggl(\frac{1+\beta^2}{2\beta}L-1\biggr)\ln\frac{1}{1-\beta^2} +F_V+F_S+F_H\biggr].
\ea
After some algebra one obtains
\ba
\Delta\sigma^{e\bar{e}\to \pi\bar{\pi}}(s)=2\frac{\alpha}{\pi}\sigma_B(s)\Delta(\beta),
\ea
with
\ba
\Delta(\beta)=-\frac{3}{2}\ln\frac{4}{1-\beta^2}-2\ln\beta+\frac{1+\beta^2}{2\beta}\biggl[-\frac{1}{12}\pi^2+\frac{5}{4}L+ \nn \\
\frac{3}{2\beta}[1-\frac{1}{2\beta}L]-L\ln\beta + Li_2(\frac{1-\beta}{1+\beta})+3Li_2(-\beta)-Li_2(\beta)+ \nn \\
3Li_2(\frac{1+\beta}{2})-2Li_2(\frac{1-\beta}{2})+2\ln\beta\ln(1+\beta)-2Li_2(1-\beta)+
\frac{1}{2}\ln^2(\frac{1+\beta}{2})\biggr].
\ea

\section{Results for FSE in point-like pion approximation}

The total contribution to $a_\mu$ can be obtained from the general formulas (see(10)) by the replacement
\ba
\beta^4 \to\beta^4[1+\frac{2\alpha}{\pi}\Delta(\beta)]=\beta^4[1+\delta(s)]
\ea
Numeric estimation leads to
\ba
\Delta^{\pi} a_\mu=-6.923\times 10^{-11}.
\ea

The total set of the lowest-order radiative corrections (RC) to $a^{(1)}_\mu$ also takes into account the
correction to the kernel
\ba
\Delta^{ker} a_\mu =\frac{\alpha^3}{3\pi^3}\int\limits_0^1\frac{\beta_\pi^4 d\beta_\pi}{1-\beta_\pi^2}K^{(2)}(b), \,\,\,\,\, b=\frac{s}{M^2}=\frac{4}{\rho^2(1-\beta_\pi^2)}.
\ea
The explicit form of the kernel $K^{(2)}(b)$ as well as its expansion in powers $b^{-1}$ are presented in the Appendix.
Using the explicit form of $K^{(2)}$ can be successfully applied to the region $1-\beta^2_\pi\sim 1$ and is not
convenient for the region $1-\beta^2_\pi<< 1$. In this region, we apply its expansion in powers of $M^2/s$, which
was obtained in \cite{Krause}. For this aim we choose an auxiliary parameter $\beta_0\sim 1$
\ba
\Delta^{ker}a_\mu=\frac{1}{4\pi^3}\biggl[\int\limits_0^{\beta_0}\frac{\beta_\pi^4 d\beta_\pi}{1-\beta_\pi^2}K^{(2)}(b)_{BR}+
\int\limits_{\beta_0}^1\frac{\beta_\pi^4 d\beta_\pi}{1-\beta_\pi^2}K^{(2)}(b)_{Kr}\biggr].
\ea
The result does not depend on $\beta_0$ and is
\ba
\Delta^{ker} a_\mu= -1.73 \cdot 10^{-10}.
\ea

The total contribution of the correction to $a_\mu$ from RC to  both the final $\pi^+\pi^-$ and the kernel is
\ba
\Delta A_\mu=\Delta^\pi+\Delta^{ker}= -2.4 \cdot 10^{-10} .
\ea

\section{Insertion of pion form factor. Discussion}
The result obtained in the pointlike approximation is about an order of magnitude lower than one measured
in experiment \cite{Davier} $a_\mu\approx 6.974\cdot 10^{-8}$.
The conversion of a virtual photon to the $\pi^+\pi^-(\gamma)$ state in the timelike region is realized through the intermediate state with vector meson $\rho(775), \omega(782), \phi(1020), \rho'(1450)$ with the following decay to the
two-pion state. The main contribution arises
from $\rho(775), \omega(782)$ meson states. Keeping in mind the resonance nature of this transition it can be taken into account by the replacement in (1)
\ba
\sigma_B(s) \to \sigma_B(s)Z,
\ea
The contribution of $\omega(782), B_\omega$  arises due to a rather small $\rho-\omega$ mixing. It has two sources:
one is connected with the quark $u, d$ mass difference $m_d-m_u=3.75 MeV, m_u=280 MeV$, and the other is connected with the transition $\omega\to\gamma \to \rho$ \cite{Volkov86} (see Fig.1)
\begin{figure}
\centering
\includegraphics[width=0.72\textwidth]{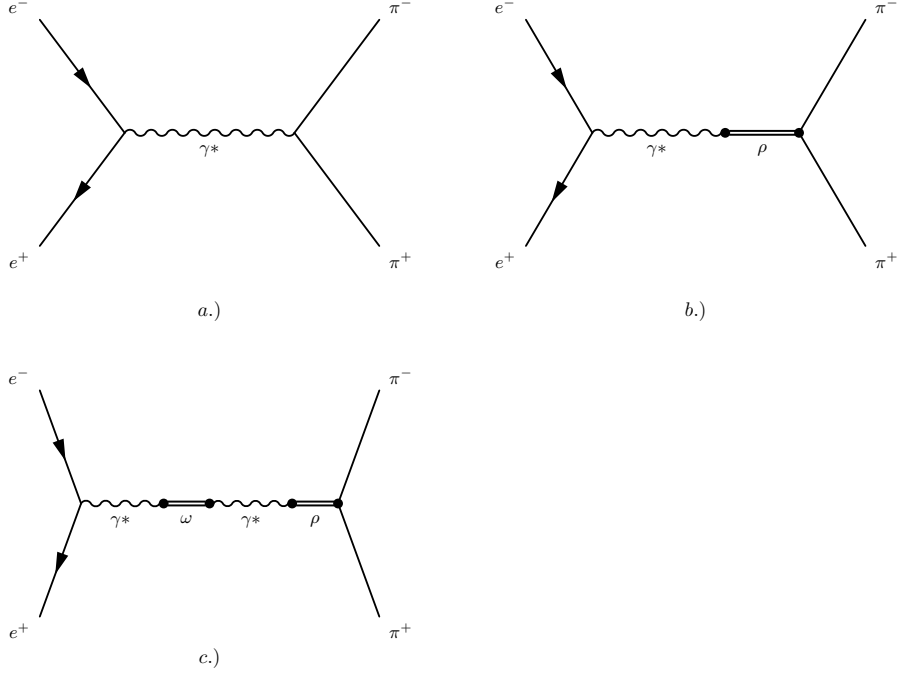}
\caption{
a) and b) NJL model approach to $B_{\gamma\rho}$; \,\,\,\,c) the same for $B_\omega$.}
\label{Fig1:1ab}
\end{figure}
\ba
B_\omega=R\frac{s}{m_\omega^2-s-i\sqrt{s}\Gamma_\omega}\frac{s}{m_\rho^2-s-i\sqrt{s}\Gamma_\rho}.
\ea
Adding the contribution of the photon and photon-rho meson conversion \cite{Volkov12}
\ba
B_{\gamma\rho}=1+\frac{s}{m_\rho^2-s-i\sqrt{s}\Gamma_\rho}=\frac{m_\rho^2-i\sqrt{s}\Gamma_\rho}{m_\rho^2-s-i\sqrt{s}\Gamma_\rho},
\ea
we have \cite{Volkov86,Volkov12}
\ba
Z(x)=|B_{\gamma\rho}+B_\omega|^2=\biggl[\frac{\mu_\rho^2-\mu_\rho+\gamma_\rho^2}{(\mu_\rho^2-1)^2+\gamma_\rho^2}+ \nn \\
R\frac{(\mu_\omega^2-1)^2-\gamma_\rho\gamma_\omega}{[(\mu_\omega^2-1)^2+\gamma_\omega^2][(\mu_\omega^2-1)^2+\gamma_\rho^2]}\biggr]^2+\nn \\
\biggl[\frac{\gamma_\rho}{(\mu_\rho^2-1)^2+\gamma_\rho^2}+ \nn \\
R\frac{(\mu_\omega^2-1)(\gamma_\rho+\gamma_\omega)}{[(\mu_\omega^2-1)^2+\gamma_\omega^2][(\mu_\omega^2-1)^2+\gamma_\rho^2]}\biggr]^2,
\ea
with \cite{Ebert}
\ba
R=\frac{1}{3g_\rho}[\frac{g_\rho^3}{16\pi^2}\ln\frac{m_d^2}{m_u^2}-\frac{4\pi\alpha}{3 g_\rho}]\approx 1.85 \cdot 10^{-3}, \nn \\
x = \frac{s}{m_{\pi}^2},\,\,\,\mu_\rho^2=\frac{m_{\rho}^2}{s}=\frac{30.86}{x}; \,\,\,
\mu_\omega^2 = \frac{m_{\omega}^2}{s}=\frac{31.43}{x}, \nn \\
g_{\rho} =6.14; \,\,\,g_{\rho}^2/(4\pi) = 3;\,\,\,\gamma_\rho^2=\frac{\Gamma_{\rho}^2}{s}=\frac{1.07}{x};\,\,\,\,
\gamma_\omega^2=\frac{\Gamma_{\omega}^2}{s}=\frac{6.08\cdot 10^{-2}}{x}.
\ea
Our final results are
\ba
a^{(1)}=\frac{\alpha^2}{12\pi^2}\int\limits_4^\infty\frac{d x}{x^2}Z(x)(1-\frac{4}{x})^{3/2} K(x),  \nn \\
\Delta a=\frac{\alpha^3}{6\pi^3}\int\limits_4^\infty\frac{d x}{x^2}Z(x)(1-\frac{4}{x})^{3/2}[\Delta(\beta)K(x)+
\Delta K(x)].
\ea
The expression for $\Delta K(x)$ is presented in the Appendix
\ba
K(x)=x K^{(1)}(x)=\int\limits_0^1\frac{y^2(1-y)x d y}{y^2+x(1-y)\rho^2}.
\ea
The explicit expression for $\Delta(\beta)$ is given above.
The result of numerical calculations is
\ba
a_{NJL}^{(1)} \approx 5.48\cdot 10^{-8};\,\,\,\,
\Delta a \approx -3.43\cdot 10^{-9}.
\ea
The contribution of the term of an order of $1/(x\rho^2)^4$  is expected to be on the level of several percent,
which determine the accuracy of our calculations.

The relative wight of $\pi^+\pi^-$ hadron state is
\ba
\frac{a_{NJL}^{(1)}}{a_{exp}} = 0.78.
\ea

Here we do not take into account the contribution of double vacuum polarization,  with the two-hadronic insertion and the QED one with the electron-positron intermediate state. Both of them were considered in the recent paper \cite{Rafael12}.

In \cite{Jeger}, an attempt to take into account the initial state emission of an additional pair of charged particles
from the experimental data was made.

In \cite{Rafael12}, a similar calculation was performed by using the duality approximation (constituent quarks and gluons and/or hadrons)
and applying the result of \cite{KS} for the final-state emission of a fermion-anti-fermion pair.

\section{Acknowledgement}
This work was supported by RFBR  Grant No. 11-02-00112. We are grateful to D.G.Kostunin for his critical reading.

\appendix
\section{}

The explicit form of  the kernel $K^{(2)}(b)$ was obtained in paper of R. Barbieri and E. Remiddi \cite{BR}.
The contribution of 14 Feynman diagram was taken into account. It has the form

\ba
K^{(2)}(b)_{BR}=-\frac{139}{144}+\frac{115}{72}b+\biggl(\frac{19}{12}-\frac{7}{36}b
+\frac{23}{144}b^2+\frac{1}{b-4}\biggr)\ln (b) + \nn \\
(-\frac{4}{3}+\frac{127}{36}b -\frac{115}{72}b^2 +\frac{23}{144}b^3)\frac{\ln y}{\sqrt{b(b-4)}}+ \nn \\
\biggl(\frac{9}{4}+\frac{5}{24}b-\frac{1}{2}b^2 -\frac{2}{b}\biggr)\xi(2)+
\frac{5}{96}b^2\ln^2 b +(-\frac{1}{2}b+\frac{17}{24}b^2 - \nn\\
\frac{7}{48}b^3)\frac{\ln y}{\sqrt{b(b-4)}}\ln b + \nn \\
\biggl(\frac{19}{24} +\frac{53}{48}b -\frac{29}{96}b^2-\frac{1}{3b} +\frac{2}{b-4}\biggr)\ln^2 y + \nn \\
(-2b +\frac{17}{6}b^2 -\frac{7}{12}b^3)\frac{1}{\sqrt{b(b-4)}}D_p(b) +
\biggl(\frac{13}{3}-\frac{7}{6}b+\frac{1}{4}b^2 - \nn \\
\frac{1}{6}b^3 -\frac{4}{b-4}\biggr)\frac{D_m(b)}{\sqrt{b(b-4)}}+ (\frac{1}{2}-\frac{7}{6}b+\frac{1}{2}b^2)T(b), \,\,\,\,\,
\ea
where
\ba
y=\frac{\sqrt{b} - \sqrt{b-4}}{\sqrt{b} + \sqrt{b-4}}, \nn \\
D_p(b) =Li_2(y) +\ln(y) \ln(1-y) -\frac{1}{4}\ln^2 y - \xi(2),  \nn \\
D_m(b) = Li_2(-y) +\frac{1}{4}\ln^2 y +\frac{1}{2}\xi(2),  \nn \\
T(b) = -6 Li_3(y) -3 Li_3(-y) +\ln^2 y \ln(1-y) + \nn \\
\frac{1}{2}(\ln^2 y + 6 \xi(2))\ln(1+y) + 2\ln y (Li_2 (-y) +2 Li_2 (y)).
\ea
The functions $Li_2(y), \,\,Li_3(y)$ are the dilogarithm and trilogarithm defined through
\ba
Li_2(y) =-\int\limits_0^y \frac{dt}{t}\ln(1-t)=-\int\limits_0^1 \frac{dt}{t}\ln(1-t y), \nn \\
Li_2(-y) =-\int\limits_0^y \frac{dt}{t}\ln(1+t)=-\int\limits_0^1 \frac{dt}{t}\ln(1+t y); \nn \\
Li_3(y) = \int\limits_0^y \frac{dt}{t}[\ln t-\ln y]\ln(1-t)= \int\limits_0^1 \frac{dt}{t}\ln t\ln(1-t y), \nn \\
Li_3(-y) = \int\limits_0^y \frac{dt}{t}[\ln t-\ln y]\ln(1+t)= \int\limits_0^1 \frac{dt}{t}\ln t\ln(1+t y).
\ea
In \cite{Krause} the expansion  $b=s/m_\mu^2$ was obtained
\ba
K^{(2)}(b)_{Kr}=\frac{1}{b}\biggl\{[\frac{223}{54}-2\xi_2-\frac{23}{36}L]+\frac{1}{b}[\frac{8785}{1152} -\frac{37}{8}\xi_2 -
\frac{367}{216}L +\frac{19}{144}L^2] + \nn \\
\frac{1}{b^2}[\frac{13072841}{432000} -\frac{883}{40}\xi_2 -\frac{10079}{3600}L +\frac{141}{80}L^2]+ \nn \\
\frac{1}{b^3 }[\frac{2034703}{16000} -\frac{3903}{40}\xi_2 -\frac{6517}{1800}L +
\frac{961}{80}L^2]\biggr\}, L=\ln b. \,\,\,\,
\ea
In the text above we use
\ba
\Delta K(x)=\frac{1}{\rho^2}[c_0+d_0L+\frac{1}{x\rho^2}[c_1+d_1L+d_2L^2]+\frac{1}{(x\rho^2)^2}[c_2+d_2L+e_2L^2]+ \nn \\
\frac{1}{(x\rho^2)^3}[c_3+c_3L+d_3L^2]], \,\,L=\ln(x\rho^2)
\ea
The numeric values are
\ba
c_0=-0.843; \,\,\,d_0=-0.639; \,\,\,e_0=0; \nn \\
c_1=0.027;\,\,\, d_1=-2.8; \,\,\,e_1=0.132; \nn \\
c_2=-6.01; \,\,\,d_2=-2.8; \,\,\,e_2=1.76; \nn \\
c_3=-33.1; \,\,\,d_3=-3.62; \,\,\,e_3=12.01.
\ea
In Fig.2 the $\beta_\pi$ dependence of the exact integrand $F_{BR}(\beta_\pi)=\beta_\pi^4K^{(2)}_{BR}/(1-\beta_\pi^2)$ and its expansion in powers of $M^2/s$ $F_{Kr}(\beta_\pi)=\beta_\pi^4K^{(2)}_{Kr}/(1-\beta_\pi^2)$ are presented.
One see the large compensations in $F_{BR}$ that take place for $\beta \to 1$.
\begin{figure}
\centering
\includegraphics[width=0.7\textwidth]{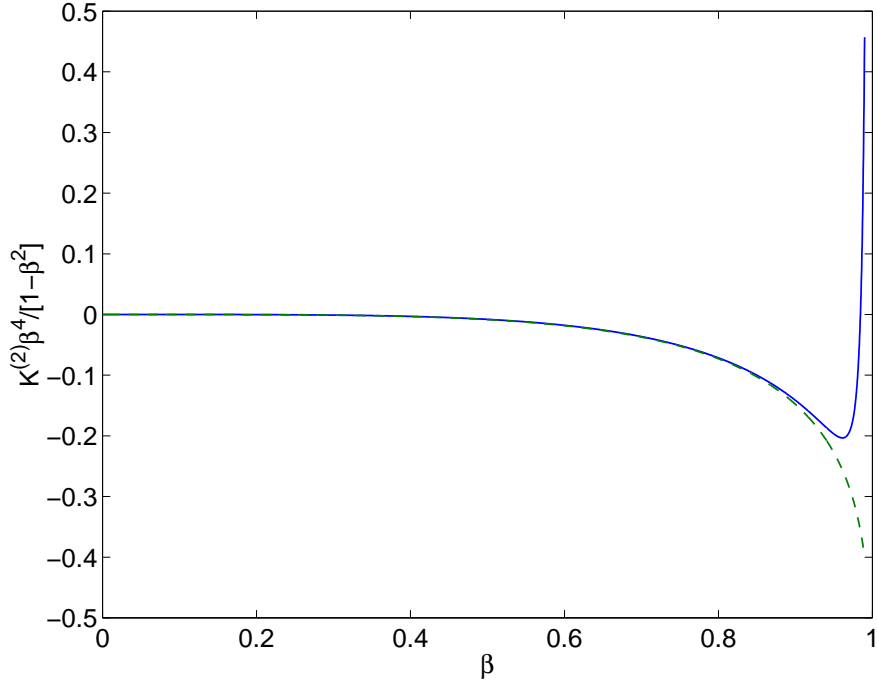}
\caption{
Dependence of $K^{(2)}(\beta)\beta^4/(1-\beta^2)$: solid line -- exact formulae,\,\,
dashed line -- power $M^2/s$ expansion.}
\label{Fig:1ab}
\end{figure}

\end{document}